\documentclass[preprint,pteplogo]{ptephy_v2}%%%%%% to generate preprint number with ptep logo

\preprintnumber{arXiv:2605.20217} %%% %%% Insert preprint number here

\usepackage{graphicx}
\usepackage{bm}
\usepackage{amsmath}
\usepackage{amssymb}
\usepackage{hyperref}
\usepackage{mathtools}
\usepackage{physics}
\usepackage{siunitx}
\begin{document}

\title{Effective Field Theory for a Baryon-Correlated Dark Matter Profile}

\author{Kento Kamada}
\affil{Department of Physics, Tohoku University, Sendai, Miyagi 980-8578, Japan \email{kento.kamada.p1@dc.tohoku.ac.jp}}

\begin{abstract}
While the standard $\Lambda$CDM model succeeds on large cosmological scales, it faces persistent small-scale challenges, including the core-cusp problem, the diversity of galaxy rotation curves, and the tight correlation between dark matter and baryons observed in the Tully-Fisher relation. To address these issues, we recently proposed an empirical law where the effective dark matter energy density is directly correlated with the baryonic gravitational potential, $\rho_{\rm DM} \propto \Phi_b^2$, which reproduces observed rotation curves and resolves the core-cusp and diversity problems. To provide a theoretical foundation for this empirical law, we construct an effective field theory (EFT) introducing massive scalar, vector, and tensor mediators between baryons and a dark sector field $\chi$. We demonstrate that aligning the mediator couplings to a specific ratio (4:6:3) with degenerate masses cancels the additional fifth forces acting on baryons up to $\mathcal{O}(v^2)$. We then show that this theoretical framework originates from a 5-dimensional (5D) spacetime. Treating the baryonic source as a 5D null fluid reveals that the three mediators emerge from a single 5D symmetric tensor field. By confining the field $\chi$ to a 4D brane, we show that its interaction with these mediators generates a pressureless energy density ($\rho_{\rm int} \propto \Phi_b^2$) that yields the empirically required baryon-correlated profile. Consequently, the field $\chi$ exhibits a scale-dependent transition: on cosmological scales, its mass energy acts as standard Cold Dark Matter (CDM), whereas on galactic scales, its interaction energy governs local dynamics. Finally, by evaluating the dynamical boundary of this localized interaction region, we provide a physical interpretation that yields the relation $\mu = K M_b^{-3/2}$, offering a theoretical basis for the Tully-Fisher relation.
\end{abstract}

\maketitle

\section{Introduction}
The standard $\Lambda$CDM model successfully describes the large-scale structure of the Universe, but faces persistent challenges on galactic scales known as the small-scale problems. These include the core-cusp problem \cite{deBlok2010, Navarro1996}, the diversity of galaxy rotation curves \cite{Oman2015}, and the tight correlation between dark matter and baryonic matter observed in the Tully-Fisher relation \cite{McGaugh2000, Lelli2016}. While particle candidates such as WIMPs \cite{Jungman1996} and Axions \cite{Preskill1983} can behave as CDM on large scales, the collisionless CDM framework faces the aforementioned small-scale issues. Conventionally, these issues are addressed by simulating baryonic feedback processes within the CDM framework \cite{Governato2010, Pontzen2012}. Other theoretical extensions propose altering the dynamics of dark matter—as seen in Self-Interacting Dark Matter (SIDM) \cite{Spergel2000} or Fuzzy Dark Matter (FDM) \cite{Hu2000} models—or modifying the laws of gravity itself (e.g., MOND \cite{Milgrom1983}, TeVeS \cite{Bekenstein2004}, and STVG \cite{Moffat2006}). In contrast to these approaches, we recently proposed an empirical law: the effective dark matter energy density $\rho_{\rm DM}$ is correlated with the baryonic gravitational potential $\Phi_b$ as $\rho_{\rm DM} = \mu \Phi_b^2 / c^4$. Because the dark matter distribution is directly governed by the baryonic potential, this model reproduces observed rotation curves using only baryonic mass profiles, resolving the core-cusp and diversity problems \cite{Kamada_empirical}.
To provide a theoretical foundation for this empirical law and investigate the physical origin of the coefficient $\mu$, we construct an effective field theory (EFT). This paper is organized as follows. In Section 2, we introduce massive scalar, vector, and tensor mediator fields, demonstrating that a specific coupling ratio (4:6:3) with degenerate masses cancels the additional fifth forces up to $\mathcal{O}(v^2)$. In Section 3, we show that this setup originates from a 5-dimensional (5D) spacetime, where the baryonic source is treated as a 5D null fluid. In Section 4, by confining a dark sector field $\chi$ to a 4D brane, we show that its interaction with the 5D mediators generates a pressureless ($T_{ij} = 0$) energy density ($\rho_{\rm int} \propto \Phi_b^2$). We then investigate the dynamics of the field $\chi$ in Section 5, establishing its scale-dependent transition from standard Cold Dark Matter (CDM) on cosmological scales to the baryon-correlated profile on galactic scales. In Section 6, by evaluating the dynamical boundary of this localized interaction region, we provide a physical interpretation yielding the relation $\mu = K M_b^{-3/2}$ for the Tully-Fisher relation. Finally, Section 7 summarizes our findings and discusses future prospects.

\section{Mediator Sector and Fifth Force Cancellation}

\subsection{Minimal Scalar Mediator Model}
To derive a dark matter energy density that correlates with the baryonic distribution, we first consider a long-range scalar field $\phi$ that couples to the trace of the macroscopic baryonic energy-momentum tensor, $\Theta_{(B)\mu}^\mu$. The Lagrangian density is given by
\begin{equation}
    \mathcal{L}_{\phi} = \frac{1}{2}\partial_\mu \phi \partial^\mu \phi - \frac{1}{2}m_\phi^2 \phi^2 + g_s \phi \Theta_{(B)\mu}^\mu.
\end{equation}
Treating the baryonic source as a pressureless dust fluid, we write its energy-momentum tensor as $\Theta_{(B)}^{\mu\nu} = \rho_b u^\mu u^\nu$, where $\rho_b$ is the Lorentz-invariant rest mass density and $u^\mu$ is the 4-velocity. Evaluating the trace yields $\Theta_{(B)\mu}^\mu = \rho_b u^\mu u_\mu = \rho_b$. Thus, the interaction term reduces to $g_s \phi \rho_b$, and the Lagrangian density becomes
\begin{equation}
    \mathcal{L}_{\phi} = \frac{1}{2}\partial_\mu \phi \partial^\mu \phi - \frac{1}{2}m_\phi^2 \phi^2 + g_s \phi \rho_b.
\end{equation}
In the static limit, the equation of motion for $\phi$ is $-\nabla^2 \phi + m_\phi^2 \phi = g_s \rho_b$. By comparing this equation of motion to the Poisson equation for the Newtonian gravitational potential ($\nabla^2 \Phi_b = 4\pi G \rho_b$), we find that the field $\phi$ traces the baryonic potential $\Phi_b(r)$ with a Yukawa suppression factor:
\begin{equation}
    \phi(r) = -\frac{g_s}{4\pi G} \Phi_b(r) e^{-m_\phi r}.
\end{equation}
The energy density of this field is defined by the $T_{00}$ component of the energy-momentum tensor as $\rho_\phi = \frac{1}{2}(\nabla\phi)^2 + \frac{1}{2}m_\phi^2 \phi^2$. Substituting the field solution, we find that the mass energy term becomes
\begin{equation}
    \frac{1}{2}m_\phi^2 \phi^2(r) = \frac{1}{2} m_\phi^2 \left( \frac{g_s}{4\pi G} \right)^2 \Phi_b^2(r) e^{-2m_\phi r}.
\end{equation}
If the mediator mass is sufficiently light such that the Yukawa suppression is negligible at galactic scales ($e^{-2m_\phi r} \simeq 1$), this mass energy provides a distribution proportional to $\Phi_b^2$.
However, this minimal approach has two fundamental issues. First, while the mass energy provides the desired $\Phi_b^2$ dependence, the gradient energy term $\frac{1}{2}(\nabla\phi)^2$, which is proportional to $(\nabla\Phi_b)^2$, introduces an extra gravitating source that does not match the empirical law. Second, the coefficient of the $\Phi_b^2$ term is a universal constant determined by fundamental parameters ($m_\phi, g_s, G$). This model cannot explain the relation $\mu = K M_b^{-3/2}$ required to satisfy the observed Tully-Fisher relation \cite{McGaugh2000}.

\subsection{Scalar Mediator and the Dark Sector Field}
To satisfy the empirical requirement for the coefficient $\mu$, we introduce a real scalar field $\chi$ in the dark sector alongside the mediator $\phi$. The Lagrangian density is given by
\begin{equation}
    \mathcal{L} = \frac{1}{2}\partial_\mu \chi \partial^\mu \chi - \frac{1}{2}m_{\chi0}^2 \chi^2 + \frac{1}{2}\partial_\mu \phi \partial^\mu \phi - \frac{1}{2}m_{\phi0}^2 \phi^2 - \frac{1}{2}g_\chi \chi^2 \phi^2 + g_s \phi \rho_b,
\end{equation}
where $m_{\chi0}$ and $m_{\phi0}$ are the bare masses of $\chi$ and $\phi$, respectively.
From this Lagrangian, the effective mass of the mediator is given by $m_{\phi, {\rm eff}}^2 = m_{\phi0}^2 + g_\chi \chi^2$. As shown in Section 2.1, if the bare mass $m_{\phi0}$ is dominant, the coefficient of the $\Phi_b^2$ term becomes a constant, which cannot explain the empirical relation. To introduce the possibility that this coefficient varies across galaxies, we assume that the interaction term is much larger than the bare mass term, $g_\chi \chi^2 \gg m_{\phi0}^2$, so that the effective mass becomes $m_{\phi, {\rm eff}}^2 \simeq g_\chi \chi^2$. As with the minimal model, this setup still introduces an extra gravitating source from the gradient energy. We will show in Section 3 that the energies of the free fields—including both the gradient and bare mass terms—vanish under a specific 5D geometric structure. Anticipating this result, we focus on the interaction energy. The effective interaction energy density is expressed as:
\begin{equation}
    \rho_{\rm int} \simeq \frac{1}{2}g_\chi \chi^2 \phi^2 \simeq \frac{1}{2}g_\chi \left(\frac{g_s}{4\pi G}\right)^2 \chi^2 \Phi_b^2 e^{-2m_{\phi, {\rm eff}}r}. \label{eq:rho_int_minimal}
\end{equation}
If the effective Compton wavelength is larger than the galactic scale, the Yukawa factor can be approximated as $e^{-2m_{\phi, {\rm eff}}r} \simeq 1$. Comparing the resulting density with the empirical law $\rho_{\rm int} = \mu \Phi_b^2/c^4$ \cite{Kamada_empirical} identifies the coefficient $\mu$ as (the derivation of $\mu \propto M_b^{-3/2}$ is detailed in Section 6):
\begin{equation}
    \mu = \frac{c^4}{2}g_\chi \left(\frac{g_s}{4\pi G}\right)^2 \chi^2. \label{eq:mu_minimal}
\end{equation}
However, this setup introduces a phenomenological problem. Because the field $\phi$ couples to the baryon density, it generates an additional attractive force between baryons, known as a fifth force. For a static test particle of mass $m$, the resulting potential is:
\begin{equation}
    V_s(r) = m \alpha_5 \Phi_b(r) e^{-m_{\phi, {\rm eff}} r} \simeq m \alpha_5 \Phi_b(r),
\end{equation}
where we define the dimensionless coupling ratio $\alpha_5 \equiv g_s^2/(4\pi G)$. To maintain the interaction energy density $\rho_{\rm int}$ at least up to the galactic scale, the effective Compton wavelength must be larger than the galaxy size ($m_{\phi, {\rm eff}}^{-1} \gtrsim R_{\rm gal}$). This requirement justifies neglecting the Yukawa suppression ($e^{-m_{\phi, {\rm eff}} r} \simeq 1$) in the potential. 
Furthermore, using the relationship $m_{\phi, {\rm eff}}^2 \simeq g_\chi \chi^2$, the coefficient $\mu$ can be rewritten as $\mu = c^4 m_{\phi, {\rm eff}}^2 \frac{\alpha_5}{8\pi G}$. Substituting the wavelength constraint ($m_{\phi, {\rm eff}} \lesssim R_{\rm gal}^{-1}$) into this equation places a lower bound on the coupling ratio:
\begin{equation}
    \alpha_5 \gtrsim 8\pi G R_{\rm gal}^2 \frac{\mu}{c^4}.
\end{equation}
To evaluate this lower bound for a typical spiral galaxy like the Milky Way, we adopt the representative baryonic mass $M_b \simeq 6.3 \times 10^{10} M_\odot$ and the observable galactic scale $R_{\rm gal} \simeq 15\, {\rm kpc}$ \cite{BlandHawthorn2016}. Using the value $\log_{10} K \simeq 70.23$ (in SI units) determined in our previous study \cite{Kamada_empirical}, the coefficient $\mu$ is calculated as $\mu = K M_b^{-3/2} \simeq 3.8 \times 10^8\, {\rm J/m^3}$. Substituting these values into the inequality yields:
\begin{equation}
    \alpha_5 \gtrsim 1.7 \times 10^7.
\end{equation}
This implies that the attractive fifth force reaches the order of $10^7$ relative to Newtonian gravity, which contradicts observed galactic dynamics. Conventionally, such large fifth forces are suppressed by invoking non-linear screening mechanisms (e.g., the Vainshtein and chameleon mechanisms \cite{Vainshtein1972, Khoury2004, Hinterbichler2012}). However, these non-linear effects distort the mediator field profiles from the linear Newtonian form ($\phi \not\propto \Phi_b$). Such distortions would destroy the proportionality $\rho_{\rm int} \propto \Phi_b^2$ on galactic scales, failing to reproduce the empirical law of dark matter energy density. In the following sections, instead of invoking non-linear screening effects, we investigate a linear mechanism to cancel this fifth force by introducing additional mediator fields, thereby preserving the desired baryon-correlated profile.
\subsection{Vector Mediator and $\mathcal{O}(v^2)$ Residual Force}
To cancel the attractive fifth force, we introduce a massive vector field $A_\mu$ that couples to the baryon current $J^\mu = \rho_b u^\mu$. As with the scalar mediator, we also introduce a coupling to the dark sector field $\chi$.  The Lagrangian density is given by
\begin{equation}
    \mathcal{L}_A = -\frac{1}{4}F_{\mu\nu}F^{\mu\nu} + \frac{1}{2}m_{A0}^2 A_\mu A^\mu + \frac{1}{2}g_\chi \chi^2 A_\mu A^\mu + g_v J^\mu A_\mu,
\end{equation}
where $F_{\mu\nu} = \partial_\mu A_\nu - \partial_\nu A_\mu$ is the field strength tensor, $m_{A0}$ is the bare mass of the vector field, and $g_v$ is the vector coupling constant. The vector field acquires an effective mass $m_{A, {\rm eff}}^2 = m_{A0}^2 + g_\chi \chi^2$.
For a static baryonic source $J^\mu = \rho_b u^\mu$ with $u^\mu = (1, \vec{0})$, the solution is $A_0(r) = \frac{g_v}{4\pi G} \Phi_b(r) e^{-m_{A, {\rm eff}} r}$ and $A_i = 0$. 
For a test particle of mass $m$, this vector field generates a repulsive Yukawa-type potential $V_v(r) = -m g_v A_0(r)$:
\begin{equation}
    V_v(r) = -m \left(\frac{g_v^2}{4\pi G}\right) \Phi_b(r) e^{-m_{A, {\rm eff}} r}.
\end{equation} 
To cancel the scalar attraction $V_s(r)$ at all distances in the static limit, the vector field must possess the equivalent coupling strength ($g_v^2 = g_s^2$) and the identical effective mass ($m_{A, {\rm eff}} = m_{\phi, {\rm eff}}$). Because both effective masses incorporate the identical interaction term $g_\chi \chi^2$, this condition demands the degeneracy of their bare masses ($m_{A0} = m_{\phi0}$). However, this static cancellation breaks down for moving test particles. In a relativistic framework, the scalar interaction is integrated over the proper time $d\tau = \gamma^{-1} dt$, introducing an inverse Lorentz factor suppression. In contrast, the vector interaction couples directly to the 4-velocity $u^\mu_{test} = \gamma(1, \vec{v}/c)$, where the $\gamma$ factor cancels this proper time suppression. Since the spatial vector components vanish ($A_i = 0$) for a static source, the effective potentials for a moving test particle become:
\begin{align}
    V_s(r, v) &= \gamma^{-1} V_s(r, 0) \simeq \left(1 - \frac{v^2}{2c^2}\right) V_s(r, 0), \\
    V_v(r, v) &= -m g_v A_0(r) = V_v(r, 0).
\end{align}
Summing these potentials under the static cancellation condition ($V_s(r,0) + V_v(r,0) = 0$) yields a residual potential at $\mathcal{O}(v^2)$:
\begin{equation}
    V_{s+v}(r, v) \simeq -m \alpha_5 \Phi_b(r) e^{-m_{\rm eff} r} \frac{v^2}{2c^2} \simeq -m \alpha_5 \Phi_b(r) \frac{v^2}{2c^2}.
\end{equation}
To estimate the effect of this residual potential, we adopt the circular velocity of the Milky Way, $v \simeq 238\, {\rm km/s}$ ($v/c \simeq 7.9 \times 10^{-4}$) \cite{BlandHawthorn2016}, as a representative galactic rotation velocity. Using the lower bound $\alpha_5 \gtrsim 1.7 \times 10^7$ derived in Section 2.2, the relative strength of this residual force compared to Newtonian gravity evaluates to $\alpha_5 (v/c)^2 / 2 \gtrsim 5$. This demonstrates that the velocity-dependent residual force is an order of magnitude stronger than Newtonian gravity. Thus, the fifth force problem is not resolved, and introducing a vector field alone is insufficient.

\subsection{Tensor Mediator and Cancellation up to $\mathcal{O}(v^2)$}
To eliminate the velocity-dependent residual force at $\mathcal{O}(v^2)$, we introduce a massive spin-2 tensor field $f_{\mu\nu}$ that couples to the baryon energy-momentum tensor $\Theta_{(B)}^{\mu\nu} = \rho_b u^\mu u^\nu$. Based on the Fierz-Pauli theory \cite{FierzPauli1939}, the dynamics of this field are governed by the following Lagrangian density:
\begin{equation}
\begin{split}
    \mathcal{L}_f &= +\frac{1}{2}\partial_\lambda f_{\mu\nu} \partial^\lambda f^{\mu\nu} - \partial_\mu f_{\nu\lambda} \partial^\nu f^{\mu\lambda} + \partial_\mu f^{\mu\nu} \partial_\nu f - \frac{1}{2}\partial_\lambda f \partial^\lambda f \\
    &\quad - \frac{1}{2}m_{f0}^2 (f_{\mu\nu}f^{\mu\nu} - f^2) - \frac{1}{2}g_\chi \chi^2 (f_{\mu\nu}f^{\mu\nu} - f^2) + g_t \Theta_{(B)}^{\mu\nu}f_{\mu\nu},
\end{split}
\end{equation}
where $f \equiv \eta^{\mu\nu}f_{\mu\nu}$ is the trace, $m_{f0}$ is the bare mass of the tensor field, and $g_t$ is the tensor coupling constant. The tensor field acquires the effective mass $m_{f, {\rm eff}}^2 = m_{f0}^2 + g_\chi \chi^2$.
To maintain the static cancellation across all distances, the three mediators must share the same bare mass ($m_{f0} = m_{A0} = m_{\phi0}$), and consequently the same effective mass $m_{\rm eff}$. This requirement of bare mass degeneracy will be resolved in Section 3, where we show they originate from a single mass parameter in a 5-dimensional spacetime.
Varying the Lagrangian for a conserved energy-momentum tensor ($\partial_\mu \Theta_{(B)}^{\mu\nu} = 0$) yields the sourced Fierz-Pauli equation \cite{Hinterbichler2012}:
\begin{equation}
(\Box + m_{\rm eff}^2) f_{\mu\nu} = g_t \left[ \Theta_{\mu\nu}^{(B)} - \frac{1}{3} \left( \eta_{\mu\nu} + \frac{\partial_\mu \partial_\nu}{m_{\rm eff}^2} \right) \Theta^{(B)} \right],
\end{equation}
where $\Theta^{(B)} \equiv \eta_{\mu\nu} \Theta_{(B)}^{\mu\nu}$ is the trace of the source.
For a static baryonic source $\Theta_{(B)}^{\mu\nu} = \rho_b u^\mu u^\nu$ with $u^\mu = (1, \vec{0})$, the static solution to the Fierz-Pauli equation is:
\begin{equation}
    f_{\mu\nu}(r) = -\frac{g_t}{4\pi G} \left( u_\mu u_\nu - \frac{1}{3}\eta_{\mu\nu} - \frac{\partial_\mu \partial_\nu}{3m_{\rm eff}^2} \right) \left[ \Phi_b(r) e^{-m_{\rm eff} r} \right].
\end{equation}
The interaction action for a test particle is $S_t = \int d^4x \, g_t f_{\mu\nu} \Theta_{\text{test}}^{\mu\nu}$. Since the test particle's energy-momentum tensor is conserved ($\partial_\mu \Theta_{\text{test}}^{\mu\nu} = 0$), the derivative term vanishes through integration by parts. Thus, only the non-derivative terms contribute to the potential. For a test particle of mass $m$ moving with velocity $v$, this generates a velocity-dependent potential:
\begin{equation}
    V_t(r, v) = m \left(\frac{g_t^2}{4\pi G}\right) \Phi_b(r) e^{-m_{\rm eff} r} \left( \gamma - \frac{1}{3}\gamma^{-1} \right).
\end{equation}
Summing the scalar, vector, and tensor potentials with the definitions $\alpha_i \equiv g_i^2/(4\pi G)$, the total potential $V_{\rm tot}$ is expressed as:
\begin{equation}
    V_{\rm tot}(r, v) = m \Phi_b(r) e^{-m_{\rm eff} r} \left[ \alpha_s \gamma^{-1} - \alpha_v + \alpha_t \left( \gamma - \frac{1}{3}\gamma^{-1} \right) \right].
\end{equation}
By expanding this expression up to $\mathcal{O}(v^2)$, we obtain:
\begin{equation}
    V_{\rm tot}(r, v) \simeq m \Phi_b(r) e^{-m_{\rm eff} r} \left[ \left( \alpha_s - \alpha_v + \frac{2}{3}\alpha_t \right) + \left( -\frac{1}{2}\alpha_s + \frac{2}{3}\alpha_t \right) \frac{v^2}{c^2} \right].
\end{equation}
To cancel the fifth force up to the $\mathcal{O}(v^2)$ term, both coefficients must independently vanish. This requirement determines the coupling constant ratio:
\begin{equation}
    g_s^2 : g_v^2 : g_t^2 = 4 : 6 : 3.
\end{equation}
Under this configuration, the static and $\mathcal{O}(v^2)$ components of the additional forces are canceled. Applying these coupling conditions yields the residual potential at the next leading order $\mathcal{O}(v^4)$ as:
\begin{equation}
    V_{\rm residual}(r, v) \simeq m \Phi_b(r) e^{-m_{\rm eff} r} \left( \frac{3}{16}\alpha_s \right) \frac{v^4}{c^4}.
\end{equation}
Using the representative values established above, the relative strength of this residual force compared to Newtonian gravity evaluates to:
\begin{equation}
    \frac{3}{16}\alpha_5(v/c)^4 \gtrsim 1.3 \times 10^{-6}.
\end{equation}
This demonstrates that the un-canceled fifth force is suppressed to the order of $10^{-6}$ relative to Newtonian gravity at the galactic scale. Thus, the fifth force is sufficiently suppressed at galactic scales, ensuring consistency with observed dynamics.
To summarize the phenomenological requirements, the combined 4D Lagrangian for the three mediators is given by:
\begin{equation}
\begin{split}
    \mathcal{L}_{\rm med}^{(4D)} &= \mathcal{L}_{{\rm kin}, \phi} + \mathcal{L}_{{\rm kin}, A} + \mathcal{L}_{{\rm kin}, f} \\
    &\quad - \frac{1}{2}m_{\phi0}^2 \phi^2 + \frac{1}{2}m_{A0}^2 A_\mu A^\mu - \frac{1}{2}m_{f0}^2 (f_{\mu\nu}f^{\mu\nu} - f^2) \\
    &\quad - \frac{1}{2}g_\chi \chi^2 \left[ \phi^2 - A_\mu A^\mu + (f_{\mu\nu}f^{\mu\nu} - f^2) \right] \\
    &\quad + g_s \rho_b \phi + g_v J^\mu A_\mu + g_t \Theta_{(B)}^{\mu\nu} f_{\mu\nu}. \label{eq:4D_combined}
\end{split}
\end{equation}
The derivation of the interaction energy density under this specific coupling configuration, and the proof that it preserves the relation $\rho_{\rm int} \propto \Phi_b^2$, are presented in Section 4, following the introduction of the 5D geometric framework in Section 3.

\section{5-Dimensional Geometric Origin and the Null Fluid Representation}
In the previous sections, the coupling ratio $g_{s}^{2}:g_{v}^{2}:g_{t}^{2}=4:6:3$ and the mass degeneracy among the bare masses of the three mediators ($m_{f0}=m_{A0}=m_{\phi0}$) were introduced to cancel the fifth force. In this section, we demonstrate that these requirements originate from a 5-dimensional geometric framework.

\subsection{5D Fierz-Pauli Action}
Inspired by the Kaluza-Klein framework \cite{Overduin1997}, we consider a 5-dimensional spacetime with the flat metric $\eta_{MN} = {\rm diag}(1, -1, -1, -1, -1)$, where the indices $M, N$ run over $0, 1, 2, 3, 5$. We denote the fifth coordinate as $y \equiv x^5$.  Assuming a compactified extra dimension, we extract the 4D effective theory by focusing on the zero mode, where the symmetric tensor field $H_{MN}$ is independent of $y$.
The dynamics of this field, coupled to the baryonic energy-momentum tensor $\mathcal{T}_{MN}^{(B)}$, are governed by the Fierz-Pauli action \cite{FierzPauli1939}. The Lagrangian density consists of three parts:
\begin{equation}
    \mathcal{L}^{(5D)} = \mathcal{L}_{\rm kin} + \mathcal{L}_{\rm mass} + \mathcal{L}_{\rm src},
\end{equation}
where the kinetic, mass, and source terms are respectively given by
\begin{align}
    \mathcal{L}_{\rm kin} &= +\frac{1}{2}\partial_P H_{MN}\partial^P H^{MN} - \partial_M H_{NP}\partial^N H^{MP} + \partial_M H^{MN}\partial_N H - \frac{1}{2}\partial_P H\partial^P H, \\
    \mathcal{L}_{\rm mass} &= -\frac{M_5^2}{2}(H_{MN}H^{MN} - H^2), \\
    \mathcal{L}_{\rm src} &= g_c \mathcal{T}^{(B)MN}H_{MN},
\end{align}
with $H \equiv \eta^{MN}H_{MN}$ and $M_5$ being the 5D bare mass.

\subsection{Dimensional Reduction of the Mass Term}
To show how the three mediator fields originate from a single 5D field, we decompose the 5D tensor $H_{MN}$ into a 4D tensor $h_{\mu\nu}$, a vector $A_\mu$, and a scalar $\phi$:
\begin{equation}
    H_{MN} = \begin{pmatrix} h_{\mu\nu} & A_\mu \\ A_\nu & \phi \end{pmatrix}.
\end{equation}
Noting that $\eta_{55} = -1$, we evaluate the scalar invariants in the 5D mass term:
\begin{align}
    H_{MN}H^{MN} &= h_{\mu\nu}h^{\mu\nu} - 2A_\mu A^\mu + \phi^2, \\
    H &= h - \phi.
\end{align}
Expanding the Fierz-Pauli mass combination $H_{MN}H^{MN} - H^2$, we obtain:
\begin{equation}
    H_{MN}H^{MN} - H^2 = h_{\mu\nu}h^{\mu\nu} - h^2 - 2A_\mu A^\mu + 2h\phi. \label{eq:mass_combination}
\end{equation}
To eliminate the mixing between the spin-2 and spin-0 degrees of freedom, we introduce the transformation $h_{\mu\nu} = f_{\mu\nu} + \frac{1}{3}\eta_{\mu\nu}\phi$, which implies the trace relation $h = f + \frac{4}{3}\phi$. Expanding each term in Eq.~(\ref{eq:mass_combination}) and summing them cancels out the cross terms $f\phi$:
\begin{equation}
    H_{MN}H^{MN} - H^2 = (f_{\mu\nu}f^{\mu\nu} - f^2) - 2A_\mu A^\mu + \frac{4}{3}\phi^2.
\end{equation}
The 4D effective mass Lagrangian $\mathcal{L}_{\rm mass}^{(4D)}$ is obtained by integrating the 5D mass Lagrangian over the extra dimension, $\mathcal{L}_{\rm mass}^{(4D)} = \int dy \mathcal{L}_{\rm mass}^{(5D)}$. Under the zero-mode approximation where the fields are independent of $y$, this integration introduces a finite length scale $L = \int dy$, which represents the size of the compactified extra dimension. Absorbing this scale into the fields to acquire the 4D mass dimension ($\hat{f}_{\mu\nu} = \sqrt{L}f_{\mu\nu}$, $\hat{A}_\mu = \sqrt{2L}A_\mu$, and $\hat{\phi} = \sqrt{4L/3}\phi$) yields the 4D mass term:
\begin{equation}
    \mathcal{L}_{\rm mass}^{(4D)} = -\frac{M_5^2}{2}(\hat{f}_{\mu\nu}\hat{f}^{\mu\nu} - \hat{f}^2) + \frac{M_5^2}{2}\hat{A}_\mu \hat{A}^\mu - \frac{M_5^2}{2}\hat{\phi}^2.
\end{equation}
These normalized fields correspond to the 4D mediators introduced in Section 2. For notational simplicity, we drop the hats hereafter ($f_{\mu\nu} \equiv \hat{f}_{\mu\nu}$, $A_\mu \equiv \hat{A}_\mu$, $\phi \equiv \hat{\phi}$). This derivation shows that the spin-2, spin-1, and spin-0 mediators are components of a single 5D tensor, inheriting the same bare mass $M_5$ and thus satisfying the requirement $m_{f0} = m_{A0} = m_{\phi0}$.

\subsection{Coupling Ratios and Null Fluid Interpretation}
We evaluate the 5D source term $\mathcal{L}_{\rm src}^{(5D)} = g_c \mathcal{T}^{(B)MN}H_{MN}$. Consider a baryon fluid moving in the 5D spacetime with a 4-velocity $u^\mu$ and a fifth velocity component $w = u^5$. The components of its energy-momentum tensor $\mathcal{T}_{MN}^{(B)}$ are:
\begin{equation}
    \mathcal{T}_{(B)}^{\mu\nu} = \rho_b u^\mu u^\nu = \Theta_{(B)}^{\mu\nu}, \quad \mathcal{T}_{(B)}^{\mu 5} = \rho_b u^\mu w = w J^\mu, \quad \mathcal{T}_{(B)}^{55} = \rho_b w^2.
\end{equation}
Expanding the interaction using the decomposed components and the transformation $h_{\mu\nu} = f_{\mu\nu} + \frac{1}{3}\eta_{\mu\nu}\phi$, we obtain:
\begin{equation}
    \mathcal{T}^{(B)MN}H_{MN} = \Theta_{(B)}^{\mu\nu}f_{\mu\nu} + 2w J^\mu A_\mu + \rho_b \left( w^2 + \frac{1}{3} \right)\phi.
\end{equation}
Applying the field normalizations from Section 3.2 (which rescales the components as $f_{\mu\nu} \to f_{\mu\nu}/\sqrt{L}$, $A_\mu \to A_\mu/\sqrt{2L}$, and $\phi \to \sqrt{3/(4L)}\phi$) and integrating over the extra dimension ($\int dy = L$), we obtain the 4D source Lagrangian $\mathcal{L}_{\rm src}^{(4D)} = \int dy \mathcal{L}_{\rm src}^{(5D)}$:
\begin{equation}
    \mathcal{L}_{\rm src}^{(4D)} = \sqrt{L} g_c \left[ \Theta_{(B)}^{\mu\nu}f_{\mu\nu} + \sqrt{2}w J^\mu A_\mu + \frac{\sqrt{3}}{2}\left( w^2 + \frac{1}{3} \right)\rho_b \phi \right].
\end{equation}
Comparing this to the combined 4D interaction in Eq.~(\ref{eq:4D_combined}) identifies the effective squared coupling constants as:
\begin{equation}
    g_t^2 = L g_c^2, \quad g_v^2 = 2w^2 L g_c^2, \quad g_s^2 = \frac{3}{4}\left( w^2 + \frac{1}{3} \right)^2 L g_c^2.
\end{equation}
The ratio of the couplings is thus determined as a function of $w$:
\begin{equation}
    g_s^2 : g_v^2 : g_t^2 = \left( \frac{9}{4}w^4 + \frac{3}{2}w^2 + \frac{1}{4} \right) : 6w^2 : 3.
\end{equation}
Equating this ratio to the requirement ($g_s^2 : g_v^2 : g_t^2 = 4 : 6 : 3$) derived in Section 2 yields a system of equations for $w^2$:
\begin{align}
    6w^2 &= 6, \\
    \frac{9}{4}w^4 + \frac{3}{2}w^2 + \frac{1}{4} &= 4.
\end{align}
Both equations are simultaneously satisfied by:
\begin{equation}
    w^2 = 1.
\end{equation}
Under this condition, the inner product of the 5D baryon velocity vector $U^M$ evaluates to:
\begin{equation}
    U^M U_M = \eta_{\mu\nu}u^\mu u^\nu + \eta_{55}(u^5)^2 = 1 - w^2 = 0.
\end{equation}
This result indicates that the baryon source behaves as a 5D null fluid. We set $w=1$ for all baryonic sources to ensure force cancellation (see Appendix \ref{sec:appendix_a} for a discussion on this sign choice). 

\subsection{Equation of Motion and Null Tensor Solution}
In the previous section, we established that the baryon fluid velocity $U^M$ is a null vector. In the static limit, the spatial velocity of the baryon fluid vanishes, and the 5D velocity reduces to a constant null vector $N^M = (1, \vec{0}, 1)$, yielding the covariant components $N_M = (1, \vec{0}, -1)$. The energy-momentum tensor of the static baryonic source is expressed as $\mathcal{T}_{AB}^{(B)} = \rho_b N_A N_B$, and its trace vanishes due to $N_A N^A = 0$:
\begin{equation}
    \mathcal{T}^{(B)} = \eta^{AB} \mathcal{T}_{AB}^{(B)} = 0.
\end{equation}
As in the 4D case in Section 2.4, the 5D equation of motion for a conserved source ($\partial^A \mathcal{T}_{AB}^{(B)} = 0$) is given by \cite{Hinterbichler2012}:
\begin{equation}
    (\Box_5 + M_5^2)H_{AB} = g_c \left[ \mathcal{T}_{AB}^{(B)} - \frac{1}{4}\left(\eta_{AB} + \frac{\partial_A \partial_B}{M_5^2}\right)\mathcal{T}^{(B)} \right].
\end{equation}
Since $\mathcal{T}^{(B)} = 0$, the equation of motion reduces to:
\begin{equation}
    (\Box_5 + M_5^2)H_{AB} = g_c \rho_b N_A N_B.
\end{equation}
Since the operator $(\Box_5 + M_5^2)$ is a scalar and the vector $N_M$ is constant, the solution $H_{AB}$ takes the same tensor structure as the source:
\begin{equation}
    H_{AB}(x) = \Psi(x) N_A N_B,
\end{equation}
where $\Psi(x)$ is a scalar amplitude. Note that this solution satisfies $H \equiv \eta^{AB}H_{AB} = \Psi (N_A N^A) = 0$. Furthermore, since the static profile $\Psi$ depends only on spatial coordinates and $N_M$ has no spatial components ($N_i = 0$), the condition $N^M \partial_M \Psi = 0$ is satisfied. Substituting this into the equation of motion yields the Klein-Gordon equation and its solution:
\begin{equation}
    (-\nabla^2 + M_5^2)\Psi = g_c \rho_b \quad \Longrightarrow \quad \Psi(r) = -\frac{g_c}{4\pi G} \Phi_b(r) e^{-M_5 r} \equiv g_c K_0(r),
\end{equation}
using the Newtonian Poisson equation $\nabla^2 \Phi_b = 4\pi G \rho_b$, where the function $K_0(r)$ defines the field profile determined by the 5D bare mass $M_5$. 
Since $N_M = (1, \vec{0}, -1)$, the non-zero components of the 5D solution are $H_{00} = H_{55} = g_c K_0(r)$ and $H_{05} = -g_c K_0(r)$. We decompose this 5D state into 4D components and isolate the spin-2 tensor via $f_{\mu\nu} = h_{\mu\nu} - \frac{1}{3}\eta_{\mu\nu}\phi$. Applying the field normalizations and the 4D coupling relations derived in Sections 3.2 and 3.3, we obtain the static profiles of the 4D mediators:
\begin{equation}
    \phi(r) = g_s K_0(r), \quad A_0(r) = -g_v K_0(r), \quad f_{00}(r) = \frac{2}{3} g_t K_0(r), \quad f_{ij}(r) = -\frac{1}{3} g_t K_0(r) \eta_{ij}.
\end{equation}
The generalization to a moving baryonic source is presented in Appendix \ref{sec:appendix_a}.

\subsection{Vanishing Free-Field Energy-Momentum Tensor}
To evaluate the energy of the mediator fields, we calculate the 5D energy-momentum tensor $T_{AB}^{(H, {\rm free})}$ for the field possessing the null structure. This energy-momentum tensor is defined by varying the action with respect to the 5D metric $G^{AB}$:
\begin{equation}
    T_{AB}^{(H, {\rm free})} = -G_{AB}\mathcal{L}_{\rm free}^{(5D)} + 2\frac{\partial \mathcal{L}_{\rm free}^{(5D)}}{\partial G^{AB}}.
\end{equation}
First, we evaluate the Lagrangian density ($\mathcal{L}_{\rm free}^{(5D)} = \mathcal{L}_{\rm kin} + \mathcal{L}_{\rm mass}$). Substituting the solution $H_{MN} = \Psi N_M N_N$ and applying the null conditions ($N_M N^M = 0$ and $H=0$), every scalar contraction vanishes:
\begin{align}
    \mathcal{L}_{\rm mass} &\propto H_{MN}H^{MN} = \Psi^2(N_M N^M)^2 = 0, \\
    \partial_P H_{MN} \partial^P H^{MN} &= (\partial_P \Psi \partial^P \Psi)(N_M N^M)^2 = 0, \\
    \partial_M H_{NP} \partial^N H^{MP} &= (\partial_M \Psi N^M)(\partial^N \Psi N_N)(N_P N^P) = 0.
\end{align}
Thus, the value of the free Lagrangian is zero ($\mathcal{L}_{\rm free}^{(5D)} = 0$), and the first term of $T_{AB}^{(H, {\rm free})}$ disappears. The energy-momentum tensor is determined by the metric derivative term $2\frac{\partial \mathcal{L}_{\rm free}^{(5D)}}{\partial G^{AB}}$ evaluated at $G=\eta$. We calculate this for each of the four kinetic terms ($L_1$ to $L_4$) and the mass term ($L_{\rm mass}$).
For the first kinetic term $L_1 = +\frac{1}{2}\partial_P H_{MN} \partial^P H^{MN}$, the variation yields:
\begin{equation}
    2\frac{\partial L_1}{\partial G^{AB}} = +\partial_A H_{MN}\partial_B H^{MN} + 2\partial_L H_{AM}\partial^L H_B^M.
\end{equation}
Substituting the null-structured solution:
\begin{align}
    +\partial_A (\Psi N_M N_N) \partial_B (\Psi N^M N^N) &= +\partial_A \Psi \partial_B \Psi (N_M N^M)^2 = 0, \\
    +2\partial_L (\Psi N_A N_M) \partial^L (\Psi N_B N^M) &= +2\partial_L \Psi \partial^L \Psi N_A N_B (N_M N^M) = 0.
\end{align}
For the second kinetic term $L_2 = -\partial_M H_{NP} \partial^N H^{MP}$, the variation yields:
\begin{equation}
    2\frac{\partial L_2}{\partial G^{AB}} = -2\partial_A H_{NP} \partial^N H_B^P - \partial_M H_{NA} \partial^N H_B^M.
\end{equation}
Substituting the solution:
\begin{align}
    -2(\partial_A \Psi N_N N_P)(\partial^N \Psi N_B N^P) &= -2\partial_A \Psi (N^N \partial_N \Psi) N_B (N_P N^P) = 0, \\
    -(\partial_M \Psi N_N N_A)(\partial^N \Psi N^M N_B) &= -(\partial_M \Psi N^M)(\partial^N \Psi N_N) N_A N_B = 0.
\end{align}
For the third kinetic term $L_3 = +\partial_M H^{MN} \partial_N H$, the variation yields:
\begin{equation}
    2\frac{\partial L_3}{\partial G^{AB}} = +2\partial_A H_B^L \partial_L H + \partial_M H^{MN} \partial_N H_{AB}.
\end{equation}
The first term vanishes since $H = 0 \Rightarrow \partial_L H = 0$. The second term becomes:
\begin{equation}
    +(\partial_M \Psi N^M N^N)(\partial_N \Psi N_A N_B) = +(N^M \partial_M \Psi)(\partial_N \Psi N^N) N_A N_B = 0.
\end{equation}
For the fourth kinetic term $L_4 = -\frac{1}{2}\partial_P H \partial^P H$, the variation gives:
\begin{equation}
    2\frac{\partial L_4}{\partial G^{AB}} = -\partial_A H \partial_B H - \partial^M H \partial_M H_{AB},
\end{equation}
which vanishes since $H = 0$.
Finally, for the mass term $L_{\rm mass} = -\frac{M_5^2}{2}(H_{MN}H^{MN} - H^2)$, the variation yields:
\begin{equation}
    2\frac{\partial L_{\rm mass}}{\partial G^{AB}} = -M_5^2(2H_{AM}H_B^M - 2H H_{AB}).
\end{equation}
Substituting the solution:
\begin{equation}
    -M_5^2(2(\Psi N_A N_M)(\Psi N_B N^M) - 0) = -2M_5^2 \Psi^2 N_A N_B (N_M N^M) = 0.
\end{equation}
Because every dummy index contraction forms either $(N_M N^M)$ or $(N^M \partial_M \Psi)$, every term vanishes under the null constraint. Consequently, the energy-momentum tensor of the free mediator field is zero:
\begin{equation}
    T_{AB}^{(H, {\rm free})} = 0.
\end{equation}
Because the free-field energy-momentum tensor is zero, the gradient and mass energies of the mediators do not gravitate. This resolves the issue discussed in Section 2, where mediators in 4D space unavoidably generate gravitating gradient and mass energies.

\section{Interaction Energy Density on a 4D Brane}
As shown in Section 3.5, the free energy of the mediator fields vanishes. If the dark sector field $\chi$ could also propagate freely in the 5D spacetime, its interaction action with the mediator fields would be defined by the 5D scalar invariant $S_{\rm int}^{\rm (5D)} \propto \int d^5X \sqrt{|G^{(5)}|} \, \chi^2(H_{MN}H^{MN}-H^2)$. The 5D energy-momentum tensor for this interaction is obtained by varying the action with respect to the full 5D metric $G^{AB}$. Just as in the free-field case, the variation of this invariant yields terms proportional to $\chi^2(H_{AM}H_B^M - H H_{AB})$. Upon substituting the null tensor ansatz $H_{MN} = \Psi N_M N_N$ and utilizing $H=0$, this reduces to $\chi^2 \Psi^2 N_A N_B (N_M N^M)$. Since $N_M N^M = 0$, both the Lagrangian value itself and its metric derivative vanish. Thus, a 5D bulk $\chi$ field yields no interaction energy density. To generate a non-vanishing interaction energy, following the braneworld framework \cite{RandallSundrum1999}, we confine the field $\chi$ to a 4D brane at $y=0$. Because $\chi$ cannot propagate in the bulk, the interaction governed by the geometric invariant discussed above is localized at the brane position, introducing a delta function $\delta(y)$ into the Lagrangian density. Furthermore, assuming the spacetime geometry takes a block-diagonal form with a dynamical 4D metric $g_{\mu\nu}$ and a fixed extra-dimensional scale $G_{55} = -1$, the 5D invariant volume measure reduces to the 4D one: $\sqrt{|G^{(5)}|} d^5X = \sqrt{-g} \, d^4x dy$.
Applying these 4D confinement conditions defines the localized interaction action as:
\begin{equation}
    S_{\rm int} = \int d^4x \int dy \sqrt{-g} \, \delta(y) \left[ -\frac{1}{2} g_{\rm 5D} \chi^2 (H_{MN}H^{MN} - H^2) \right].
\end{equation}
Integration over the fifth coordinate $y$ via the delta function evaluates the integrand at $y=0$. By substituting the rescaled 4D fields from Section 3.2 (e.g., $f_{\mu\nu} \to f_{\mu\nu}/\sqrt{L}$), the invariant $(H_{MN}H^{MN} - H^2)$ yields a common factor of $1/L$. Absorbing this scale into the 5D coupling defines the effective dimensionless 4D coupling $g_\chi \equiv g_{\rm 5D}/L$. Thus, the interaction Lagrangian on the 4D brane becomes:
\begin{equation}
    \mathcal{L}_{\chi, {\rm int}} = -\frac{1}{2}g_\chi \chi^2 \left[ \phi^2 - g^{\mu\nu}A_\mu A_\nu + g^{\mu\alpha}g^{\nu\beta}f_{\mu\nu}f_{\alpha\beta} - (g^{\mu\nu}f_{\mu\nu})^2 \right] \label{eq:L_int_def}.
\end{equation}
Since the field $\chi$ is confined to the 4D brane, its energy-momentum tensor $T_{\mu\nu}$ acting as a source for 4D gravity is obtained by varying the action with respect to the dynamical 4D metric $g^{\mu\nu}$:
\begin{equation}
    T_{\mu\nu} = \frac{2}{\sqrt{-g}}\frac{\delta(\sqrt{-g}\mathcal{L}_{\chi, {\rm int}})}{\delta g^{\mu\nu}} = -g_{\mu\nu}\mathcal{L}_{\chi, {\rm int}} + 2\frac{\partial \mathcal{L}_{\chi, {\rm int}}}{\partial g^{\mu\nu}} \label{eq:T_mu_nu_var}.
\end{equation}
First, we determine the value of the Lagrangian $-\mathcal{L}_{\chi, {\rm int}}$ itself. Notice that the interaction Lagrangian in Eq.~(\ref{eq:L_int_def}) provides the additional mass terms for all the 4D mediators. Due to this interaction, the mediators acquire the effective mass $m_{\rm eff} = \sqrt{M_5^2 + g_\chi \chi^2}$. As a result, the static field solutions generated by the baryon source are modified from the profile $K_0(r)$ to the profile $K(r) \equiv -\frac{\Phi_b(r)}{4\pi G} e^{-m_{\rm eff} r}$. Substituting these modified components ($\phi = g_s K(r)$, $A_0 = -g_v K(r)$, $f_{00} = \frac{2}{3}g_t K(r)$, and $f_{ij} = -\frac{1}{3}g_t K(r)\eta_{ij}$), the metric contraction for the vector field yields $g^{\mu\nu}A_\mu A_\nu = A_0^2 = g_v^2 K(r)^2$. For the tensor field, utilizing the trace $f = -\frac{1}{3}g_t K(r)$, the scalar invariant evaluates to:
\begin{equation}
    f_{\mu\nu}f^{\mu\nu} - f^2 = \left( f_{00}^2 + f_{ij}f^{ij} \right) - f^2 = \left( \frac{4}{9} + \frac{3}{9} \right) g_t^2 K(r)^2 - \frac{1}{9} g_t^2 K(r)^2 = \frac{2}{3} g_t^2 K(r)^2.
\end{equation}
Substituting these evaluated scalar combinations back into the interaction Lagrangian, we obtain:
\begin{equation}
    \mathcal{L}_{\chi, {\rm int}} = -\frac{1}{2}g_\chi \chi^2 \left( g_s^2 - g_v^2 + \frac{2}{3}g_t^2 \right) K(r)^2.
\end{equation}
Under the static force cancellation condition derived in Section 2 ($g_s^2 - g_v^2 + \frac{2}{3}g_t^2 = 0$), this interaction Lagrangian vanishes ($\mathcal{L}_{\chi, {\rm int}} = 0$).
Therefore, the energy density $T_{00}$ is determined by the metric derivative $2\frac{\partial \mathcal{L}_{\chi, {\rm int}}}{\partial g^{00}}$. Since the scalar field term $\phi^2$ does not depend on the metric $g^{\mu\nu}$, it does not contribute to this derivative. Thus, only the vector and tensor components yield non-vanishing terms:
\begin{equation}
    T_{00} = 2\frac{\partial \mathcal{L}_{\chi, {\rm int}}}{\partial g^{00}} = g_\chi \chi^2 \left[ A_0^2 - 2f_{0\lambda}f_0^\lambda + 2f f_{00} \right].
\end{equation}
Substituting the static components $A_0 = -g_v K(r)$, $f_{00} = \frac{2}{3}g_t K(r)$, and $f = -\frac{1}{3}g_t K(r)$ yields:
\begin{equation}
    T_{00} = g_\chi \chi^2 \left( g_v^2 - \frac{4}{3}g_t^2 \right) K(r)^2.
\end{equation}
Applying the conditions $g_v^2 = 2g_t^2$ and $g_s^2 = \frac{4}{3}g_t^2$ derived in Section 2, the scalar coefficient evaluates to $\left(g_v^2 - \frac{4}{3}g_t^2\right) = \frac{2}{3}g_t^2 = \frac{1}{2}g_s^2$. Substituting $K(r)$, the energy density becomes:
\begin{equation}
    T_{00} = \left[ \frac{1}{2} \frac{g_\chi g_s^2}{(4\pi G)^2} \chi^2 \right] \Phi_b^2(r) e^{-2m_{\rm eff} r}. \label{eq:T00_gs}
\end{equation}
This result reproduces the interaction energy density of the minimal scalar model assumed in Section 2.2. The 5D framework cancels the fifth force while retaining the scalar energy density required to satisfy the empirical law.
Provided that the effective Compton wavelength is sufficiently larger than the galactic scale ($m_{\rm eff}^{-1} \gtrsim R_{\rm gal}$), the Yukawa decay is negligible ($e^{-2m_{\rm eff} r} \simeq 1$), and the energy density reduces to $T_{00} \propto \chi^2 \Phi_b^2$.
The pressure corresponds to the spatial components $T_{ij}$ of the energy-momentum tensor:
\begin{equation}
    T_{ij} = 2\frac{\partial \mathcal{L}_{\chi, {\rm int}}}{\partial g^{ij}} = g_\chi \chi^2 \left[ A_i A_j - 2f_{i\lambda}f_j^\lambda + 2f f_{ij} \right].
\end{equation}
In the static limit, $A_i = 0$ and $f_{ij} = -\frac{1}{3}g_t K(r) \eta_{ij}$. The tensor contraction terms evaluate to:
\begin{align}
    -2f_{i\lambda}f_j^\lambda &= -2 \left(-\frac{1}{3}g_t K(r) \eta_{i\alpha}\right) \left(-\frac{1}{3}g_t K(r) \eta_{j\beta}\right) \eta^{\alpha\beta} = -\frac{2}{9}g_t^2 K(r)^2 \eta_{ij}, \\
    2f f_{ij} &= 2 \left(-\frac{1}{3}g_t K(r)\right) \left(-\frac{1}{3}g_t K(r) \eta_{ij}\right) = \frac{2}{9}g_t^2 K(r)^2 \eta_{ij}.
\end{align}
These terms cancel each other, yielding $T_{ij} = 0$. This confirms that the interaction energy behaves as a pressureless dust fluid (positive energy density and zero pressure) whose density is proportional to the square of the baryonic gravitational potential. The verification of the spatial stress components for a moving source is given in Appendix \ref{sec:appendix_a}.

\section{Dynamics of the Field $\chi$}

\subsection{Mass Shift Cancellation and Pressureless Behavior of $\chi$}

While Section 4 established that the interaction sector yields the energy density required by the empirical law, we must also investigate the local dynamics of the field $\chi$ itself. Inside a galaxy, where the effect of cosmic expansion can be neglected, the concentrated baryons generate the mediator fields, which couple to $\chi$. To determine how this baryonic environment affects the effective mass and the spatial profile of $\chi$, we evaluate its total Lagrangian density:
\begin{equation}
    \mathcal{L}_\chi = \frac{1}{2}\partial_\mu \chi \partial^\mu \chi - \frac{1}{2}m_{\chi0}^2 \chi^2 + \mathcal{L}_{\chi, {\rm int}},
\end{equation}
where $m_{\chi0}$ is the bare mass. The interactions with the mediator fields modify the effective mass of $\chi$ through a spatially dependent mass shift $\Delta m^2(x)$:
\begin{equation}
    m_{\rm eff, \chi}^2(x) = m_{\chi0}^2 + \Delta m^2(x).
\end{equation}
Based on the form of $\mathcal{L}_{\chi, {\rm int}}$, this mass shift is identified as:
\begin{equation}
    \Delta m^2(x) = g_\chi \left[ \phi^2 - g^{\mu\nu}A_\mu A_\nu + g^{\mu\alpha}g^{\nu\beta}f_{\mu\nu}f_{\alpha\beta} - (g^{\mu\nu}f_{\mu\nu})^2 \right].
\end{equation}
As shown in Section 4, this combination vanishes everywhere in the static limit due to the coupling condition $g_s^2 - g_v^2 + \frac{2}{3}g_t^2 = 0$. Thus, the mass shift is zero:
\begin{equation}
    \Delta m^2(x) = 0 \quad \Longrightarrow \quad m_{\rm eff, \chi}^2(x) = m_{\chi0}^2.
\end{equation}
Because this mass shift vanishes, the equation of motion for the field $\chi$ is simply:
\begin{equation}
    \left( \partial_\mu \partial^\mu + m_{\chi0}^2 \right) \chi = 0.
\end{equation}
This result shows that even when the field $\chi$ coexists with baryons inside a galaxy, its effective mass remains unchanged. Therefore, the field obeys the free Klein-Gordon equation as it does in a vacuum.
If we consider the case where the field $\chi$ is non-relativistic, it varies slowly in space ($|\nabla \chi| \ll m_{\chi0}\chi$). By neglecting the spatial gradients, the equation of motion reduces to:
\begin{equation}
    \ddot{\chi} + m_{\chi0}^2 \chi = 0.
\end{equation}
This yields a spatially uniform oscillation:
\begin{equation}
    \chi(t) = \chi_0 \cos(m_{\chi0} t + \delta).
\end{equation}
The energy density and pressure for this oscillating field are given by $\rho_\chi = \frac{1}{2}\dot{\chi}^2 + \frac{1}{2}m_{\chi0}^2 \chi^2$ and $P_\chi = \frac{1}{2}\dot{\chi}^2 - \frac{1}{2}m_{\chi0}^2 \chi^2$. Assuming the oscillation period is much shorter than the typical dynamical timescale of the galaxy, we take the time average over an oscillation period to obtain:
\begin{equation}
    \langle \rho_\chi \rangle = \frac{1}{2}m_{\chi0}^2 \chi_0^2, \quad \langle P_\chi \rangle = 0.
\end{equation}
These results confirm that as long as the field $\chi$ is non-relativistic, it behaves as a pressureless dust fluid whether inside a baryonic region or in a vacuum.

\subsection{Time-Averaged Interaction Energy Density}
In Section 4, we derived the energy density of the interaction sector:
\begin{equation}
    T_{00} = \left[ \frac{1}{2} \frac{g_\chi g_s^2}{(4\pi G)^2} \chi^2 \right] \Phi_b^2(r) e^{-2m_{\rm eff} r}.
\end{equation}
Because of the rapid oscillation of the field $\chi$, the energy density $T_{00}$ also fluctuates. Averaging over an oscillation period with $\langle \chi^2 \rangle = \frac{1}{2}\chi_0^2$, the time-averaged interaction energy density $\rho_{\rm int} \equiv \langle T_{00} \rangle$ is:
\begin{equation}
    \rho_{\rm int} = \left[ \frac{1}{4} \frac{g_\chi g_s^2}{(4\pi G)^2} \chi_0^2 \right] \Phi_b^2(r) e^{-2m_{\rm eff} r}.
\end{equation}
As discussed in Section 2, the effective Compton wavelength is larger than the galactic scale, allowing us to approximate the Yukawa factor as $e^{-2m_{\rm eff} r} \simeq 1$. Thus, the interaction energy density simplifies to:
\begin{equation}
    \rho_{\rm int} \simeq \left[ \frac{1}{4} \frac{g_\chi g_s^2}{(4\pi G)^2} \chi_0^2 \right] \Phi_b^2(r).
\end{equation}
Comparing this result with the empirical law $\rho_{\rm int} = \mu \Phi_b^2 / c^4$, the coefficient $\mu$ is identified as:
\begin{equation}
    \mu = \frac{c^4}{4} \frac{g_\chi g_s^2}{(4\pi G)^2} \chi_0^2 = \frac{c^4 g_\chi g_s^2}{64\pi^2 G^2} \chi_0^2. \label{eq:mu_def}
\end{equation}

\subsection{Scale-Dependent Role of the Dark Sector}

On cosmological scales, the field $\chi$ evolves in the Friedmann-Lema\^{i}tre-Robertson-Walker (FLRW) background. As established in Section 5.1, the cancellation of the mass shift ensures that $\chi$ obeys the free equation of motion regardless of the cosmological baryon density. Neglecting spatial gradients, the equation of motion is given by:
\begin{equation}
    \ddot{\chi} + 3H\dot{\chi} + m_{\chi0}^2 \chi = 0,
\end{equation}
where $H$ is the Hubble parameter. Once the expansion rate drops below the bare mass ($H \ll m_{\chi0}$), the field oscillates. Applying the WKB approximation $\chi(t) = \chi_0(t) \cos(m_{\chi0} t)$ and averaging over an oscillation period, the amplitude is governed by $2\dot{\chi}_0 + 3H\chi_0 = 0$, which yields $\chi_0(t) \propto a(t)^{-3/2}$ \cite{Marsh2016}. Consequently, the time-averaged mass energy density dilutes as $\langle \rho_\chi \rangle = \frac{1}{2}m_{\chi0}^2 \chi_0(t)^2 \propto a^{-3}$. Since the time-averaged pressure vanishes ($\langle P_\chi \rangle = 0$), the mass energy of $\chi$ acts as cold dark matter (CDM), forming halos that provide the gravitational seeds for baryons.
Once baryons fall into these potential wells, they condense to form galaxies and generate the mediator fields, which couple to the pre-existing $\chi$ halo. Provided $m_{\chi0}$ is sufficiently small, this interaction energy dominates over the mass energy of $\chi$ within the galactic region. As established in Sections 4 and 5.2, this pressureless interaction energy yields the empirical density profile $\rho_{\rm int} = \frac{\mu}{c^4} \Phi_b^2$ \cite{Kamada_empirical}, which maintains the observed galaxy rotation curves. Thus, the primary gravitational source of the dark sector shifts from the mass energy on cosmic scales to the interaction energy within galaxies.

\section{Interpretation of the Relation $\mu \propto M_b^{-3/2}$}

Having established the properties of the mediator fields and the field $\chi$ from the 5D geometric framework, we now investigate the physical origin of the relation $\mu \propto M_b^{-3/2}$. As discussed in Section 5.3, the field $\chi$ transitions from standard Cold Dark Matter on cosmological scales to a localized interaction mode driven by condensed baryons. By determining the physical boundary $R$ of this localized region and evaluating the field properties within it, we derive how the coefficient $\mu$ relates to the galactic baryonic mass $M_b$.
In a galactic system, we define this physical boundary $R$ as the radius where the baryonic gravitational acceleration $a_b$ and the acceleration from the interaction energy $a_{\rm int}$ balance each other. At the outer regions, the baryonic acceleration $a_b(R) \simeq G M_b / R^2$ is insufficient to maintain the flat rotation curves. Therefore, the interaction component must become the dominant gravitational source beyond this scale.
To determine this radius $R$, we first evaluate the effective mass of the interaction component enclosed within $R$. Since the interaction energy density is $\rho_{\rm int} = \frac{\mu}{c^4}\Phi_b^2(x)$, the corresponding mass density is $\rho_{\rm mass} = \frac{\mu}{c^6}\Phi_b^2(x)$. By integrating this mass density, we obtain:
\begin{equation}
    M_{\rm int}(R) = \int_{V_R} \frac{\mu}{c^6} \Phi_b^2(x) d^3x.
\end{equation}
From dimensional analysis, this integral is proportional to $\mu G^2 M_b^2 R / c^6$. Using a dimensionless shape factor $\xi \sim \mathcal{O}(1)$ to represent the geometry of the baryons, the mass is:
\begin{equation}
    M_{\rm int}(R) = \xi \frac{\mu G^2 M_b^2}{c^6} R.
\end{equation}
The corresponding acceleration is $a_{\rm int}(R) = G M_{\rm int}(R) / R^2$. Equating this to the baryonic acceleration ($a_{\rm int} = a_b$) yields the physical boundary $R$:
\begin{equation}
    \frac{G M_b}{R^2} = \xi \frac{\mu G^3 M_b^2}{c^6 R} \quad \Longrightarrow \quad R = \frac{c^6}{\xi G^2 M_b \mu}. \label{eq:R_dynamic}
\end{equation}
Within this determined boundary $R$, the field $\chi$ settles into the localized mode. As shown in Section 5, the cancellation of the mass shift ensures that the field obeys the free equation of motion ($\ddot{\chi} + m_{\chi0}^2\chi = 0$), oscillating rapidly in time while maintaining a uniform spatial amplitude $\chi_0$ inside $R$. We assume that the total integrated square of this amplitude over the volume $V_R$ is fixed to a constant $C_\chi$:
\begin{equation}
    \int_{V_R} \chi_0^2 d^3x = C_\chi.
\end{equation}
Since $\chi_0$ is constant within $R$, this gives $\chi_0^2 = \frac{3C_\chi}{4\pi R^3}$. Substituting this into the expression for the coefficient $\mu = \frac{c^4 g_\chi g_s^2}{64\pi^2 G^2} \chi_0^2$ from Section 5.2, we obtain:
\begin{equation}
    \mu = \frac{3 c^4 g_\chi g_s^2 C_\chi}{256\pi^3 G^2 R^3}.
\end{equation}
Finally, by substituting the radius $R$ from Eq.~(\ref{eq:R_dynamic}), we eliminate $R$:
\begin{equation}
    \mu = \left( \frac{3 c^4 g_\chi g_s^2 C_\chi}{256\pi^3 G^2} \right) \left( \frac{\xi G^2 M_b \mu}{c^6} \right)^3 = \frac{3 g_\chi g_s^2 C_\chi \xi^3 G^4 M_b^3 \mu^3}{256\pi^3 c^{14}}.
\end{equation}
Solving for the coefficient $\mu$, we arrive at the final expression:
\begin{equation}
    \mu = \left( \frac{c^7}{G^2} \sqrt{\frac{256\pi^3}{3 g_\chi g_s^2 C_\chi \xi^3}} \right) M_b^{-3/2} \equiv K M_b^{-3/2}.
\end{equation}
By comparing this theoretical expression for the parameter $K$ with the value obtained from galactic observations, we can place direct constraints on the parameters ($g_\chi$, $g_s$, and $C_\chi$).
It should be noted that the physical origin of the constant $C_\chi$ is not addressed within the framework of this low-energy effective field theory. Investigating the fundamental mechanism that dynamically fixes this integrated value requires extending the current framework to a higher-energy complete theory, which remains a subject for future work.

\section{Summary and Prospects}
In this paper, we constructed an effective field theory (EFT) to provide a theoretical foundation for the empirical law of the baryon-correlated dark matter energy density, $\rho_{\rm DM} = \mu \Phi_b^2 / c^4$ \cite{Kamada_empirical}. By introducing massive scalar, vector, and tensor fields that couple to baryons with a coupling ratio of 4:6:3 and degenerate masses, the fifth force between baryons is canceled up to $\mathcal{O}(v^2)$. We demonstrated that this EFT originates from a 5-dimensional (5D) geometry. Treating the baryonic source as a 5D null fluid reveals that the three mediators emerge from a single 5D symmetric tensor field, which dictates the required coupling ratio and mass degeneracy.
Furthermore, we showed that by confining the dark sector field $\chi$ to a 4D brane, its interaction with these mediators generates a pressureless ($T_{ij} = 0$) energy density distribution that reproduces the empirically required form $\rho_{\rm int} \propto \Phi_b^2$. In this framework, the mass energy of the field $\chi$ accounts for standard cold dark matter on large cosmological scales, while its interaction energy governs galactic dynamics. Finally, by evaluating the dynamical boundary of this localized interaction region, we provided a physical interpretation yielding the relation $\mu = K M_b^{-3/2}$, offering a theoretical basis for the origin of the Tully-Fisher relation.
Several key areas remain for future exploration. First, the current minimal model assumes that the 5D mediator $H_{MN}$ couples to baryons. Whether it also couples to other Standard Model particles remains an open question. Second, while the residual fifth force is suppressed to the order of $10^{-6}$ relative to Newtonian gravity at the galactic scale, the cancellation relies on the non-relativistic approximation. At relativistic velocities, the $\mathcal{O}(v^4/c^4)$ residual force becomes significant. Addressing this requires finding a mechanism that cancels the fifth force to all orders in $v/c$. Third, extending this model to intermediate cluster scales, such as merging clusters, requires further investigation. Determining the dynamics of the field $\chi$ and the interacting gas in these non-equilibrium environments is left for future work. Fourth, the cosmological evolution of the field $\chi$ must be clarified. Investigating its thermal history and production mechanism in the early Universe \cite{KolbTurner1990}, and verifying its consistency with Cosmic Microwave Background (CMB) data \cite{Planck2018} are next steps. Fifth, to clarify the physical origins of both the 5D null fluid motion ($w^2=1$) of baryons and the confinement of the field $\chi$ to a 4D brane, this effective field theory must be embedded into a complete ultraviolet (UV) framework. Finally, the existence of such an ultralight scalar field $\chi$ is well-aligned with the Axion-Like Particles (ALPs) predicted by the String Axiverse \cite{Arvanitaki2010}.

\appendix
\section{Kinematic Generalization of the Baryonic Source}
\label{sec:appendix_a}

\subsection{Mediator Profiles for a Moving Baryonic Source}
When the baryonic source moves with a spatially varying 4-velocity $u^\mu(\vec{r}) = \gamma(\vec{r})(1, \vec{v}(\vec{r})/c)$, the 5D null fluid condition ($w=1$) established in Section 3 gives the contravariant 5D velocity $U^M(\vec{r}) = (u^\mu(\vec{r}), 1)$, yielding the covariant components $U_M(\vec{r}) = (u_\mu(\vec{r}), -1)$.
The energy-momentum tensor of the source is $\mathcal{T}_{MN}^{(B)}(\vec{r}) = \rho_b(\vec{r}) U_M(\vec{r}) U_N(\vec{r})$. The static solution is obtained by integrating the source with the corresponding Yukawa Green's function $G(\vec{r}, \vec{r}\,')$:
\begin{equation}
    H_{MN}(\vec{r}) = g_c \int G(\vec{r}, \vec{r}\,') \rho_b(\vec{r}\,') U_M(\vec{r}\,') U_N(\vec{r}\,') d^3r'.
\end{equation}
In the non-relativistic limit ($v \ll c$, $\gamma \simeq 1$), the covariant velocity components are approximated as $U_M \simeq (1, -v_i/c, -1)$. The spatial integration of the density yields the scalar profile $K(\vec{r})$ introduced in Section 4. We define the vector potential $V_i(\vec{r})$ as:
\begin{equation}
    K(\vec{r}) = \int G(\vec{r}, \vec{r}\,') \rho_b(\vec{r}\,') d^3r', \quad V_i(\vec{r}) \equiv \int G(\vec{r}, \vec{r}\,') \rho_b(\vec{r}\,') \frac{v_i(\vec{r}\,')}{c} d^3r'.
\end{equation}
Substituting the velocity components into the source tensor up to $\mathcal{O}(v)$, the components of the 5D field $H_{MN}$ evaluate to:
\begin{equation}
    H_{00} \simeq g_c K, \quad H_{0i} \simeq -g_c V_i, \quad H_{ij} \simeq 0, \quad H_{05} \simeq -g_c K, \quad H_{i5} \simeq g_c V_i, \quad H_{55} \simeq g_c K.
\end{equation}
Following the decomposition defined in Section 3.2, $H_{MN}$ is separated into the 4D tensor $h_{\mu\nu}$, vector $A_\mu$, and scalar $\phi$. Identifying these components yields:
\begin{equation}
    h_{00} \simeq g_c K, \quad h_{0i} \simeq -g_c V_i, \quad h_{ij} \simeq 0, \quad A_0 \simeq -g_c K, \quad A_i \simeq g_c V_i, \quad \phi \simeq g_c K.
\end{equation}
Next, applying the transformation $f_{\mu\nu} = h_{\mu\nu} - \frac{1}{3}\eta_{\mu\nu}\phi$ to isolate the spin-2 degrees of freedom yields:
\begin{equation}
    f_{00} \simeq \frac{2}{3}g_c K, \quad f_{0i} \simeq -g_c V_i, \quad f_{ij} \simeq -\frac{1}{3}g_c K \eta_{ij}.
\end{equation}
Finally, we extract the physical 4D mediator fields by applying the field normalizations from Section 3.2 (which rescales the components as $f_{\mu\nu} \to f_{\mu\nu}/\sqrt{L}$, $A_\mu \to A_\mu/\sqrt{2L}$, and $\phi \to \sqrt{3/(4L)}\phi$). Utilizing the coupling constant relations ($g_t = \sqrt{L}g_c$, $g_v = \sqrt{2L}g_c$, $g_s = \sqrt{4L/3}g_c$), we obtain the fields for the moving source:
\begin{equation}
    \phi(\vec{r}) = g_s K(\vec{r}), \quad A_0(\vec{r}) = -g_v K(\vec{r}), \quad A_i(\vec{r}) = g_v V_i(\vec{r}),
\end{equation}
\begin{equation}
    f_{00}(\vec{r}) \simeq \frac{2}{3} g_t K(\vec{r}), \quad f_{0i}(\vec{r}) \simeq -g_t V_i(\vec{r}), \quad f_{ij}(\vec{r}) \simeq -\frac{1}{3} g_t K(\vec{r}) \eta_{ij}.
\end{equation}

\subsection{Fifth Force Cancellation for Moving Sources}
To evaluate the potential of the fifth force mediated by the 5D tensor field $H_{MN}$, we consider the interaction between baryonic sources. For a test particle of mass $m$ moving with a 5-velocity $U_{\rm test}^M = (u_{\rm test}^\mu, 1)$, the 4D effective interaction potential $V_{\rm tot}$ is obtained by integrating the 5D interaction over the fifth coordinate ($\int dy = L$):
\begin{equation}
    V_{\rm tot}(\vec{r}) = \int dy \left( -m g_c H_{MN} U_{\rm test}^M U_{\rm test}^N \right) = -m (L g_c) H_{MN}(\vec{r}) U_{\rm test}^M U_{\rm test}^N.
\end{equation}
Substituting the solution for $H_{MN}$ derived in Appendix A.1, which is proportional to the 5D coupling $g_c$, we obtain:
\begin{equation}
    V_{\rm tot}(\vec{r}) = -m (L g_c^2) \int G(\vec{r}, \vec{r}\,') \rho_b(\vec{r}\,') \left[ U_{\rm src}(\vec{r}\,') \cdot U_{\rm test}(\vec{r}) \right]^2 d^3r',
\end{equation}
where $U_{\rm src} \cdot U_{\rm test} \equiv \eta_{MN} U_{\rm src}^M(\vec{r}\,') U_{\rm test}^N(\vec{r})$. Applying the coupling relation $g_t^2 = L g_c^2$ derived in Section 3.3, the coefficient reduces to the 4D tensor coupling $-m g_t^2$. Since $w=1$ is set for all baryonic sources in Section 3\footnote{If interacting baryons possessed opposite fifth velocity components ($w_{\rm src} w_{\rm test} = -1$), the inner product would yield $U_{\rm src} \cdot U_{\rm test} \simeq 2$ in the non-relativistic limit, which generates a huge un-canceled fifth force.}, this 5D invariant inner product evaluates to $U_{\rm src} \cdot U_{\rm test} = \eta_{\mu\nu} u_{\rm src}^\mu u_{\rm test}^\nu - 1$.
Using the 4-velocities defined in Appendix A.1, this 4D Lorentz inner product in the non-relativistic limit ($\gamma \simeq 1 + v^2/2c^2$) expands up to $\mathcal{O}(v^2)$ as:
\begin{equation}
\begin{split}
    \eta_{\mu\nu} u_{\rm src}^\mu u_{\rm test}^\nu &= \gamma_{\rm src} \gamma_{\rm test} \left( 1 - \frac{\vec{v}_{\rm src} \cdot \vec{v}_{\rm test}}{c^2} \right) \\
    &\simeq 1 + \frac{v_{\rm src}^2 - 2\vec{v}_{\rm src} \cdot \vec{v}_{\rm test} + v_{\rm test}^2}{2c^2} \equiv 1 + \Delta,
\end{split}
\end{equation}
where $\Delta \equiv |\vec{v}_{\rm test} - \vec{v}_{\rm src}|^2 / 2c^2$ characterizes the relative velocity dependence. Substituting this result back into the 5D inner product gives:
\begin{equation}
    U_{\rm src} \cdot U_{\rm test} \simeq (1 + \Delta) - 1 = \Delta.
\end{equation}
Consequently, the total potential reduces to:
\begin{equation}
    V_{\rm tot}(\vec{r}) \simeq -m g_t^2 \int G(\vec{r}, \vec{r}\,') \rho_b(\vec{r}\,') \Delta^2 d^3r'.
\end{equation}
Since galactic baryons rotate with a typical velocity $v_{\rm typ}$ (e.g., $v \simeq 238~{\rm km/s}$ as adopted in Section 2.3), their maximum relative velocity is $|\vec{v}_{\rm test} - \vec{v}_{\rm src}| \sim \mathcal{O}(v_{\rm typ})$. This restricts the kinematic factor to $\Delta^2 \lesssim \mathcal{O}(v_{\rm typ}^4/c^4)$. Using the scalar profile $K(\vec{r})$ from Section 4, the total potential satisfies:
\begin{equation}
    |V_{\rm tot}(\vec{r})| \lesssim m g_t^2 |K(\vec{r})| \mathcal{O}\left(\frac{v_{\rm typ}^4}{c^4}\right).
\end{equation}
Substituting $K(\vec{r}) = -\frac{\Phi_b(\vec{r})}{4\pi G} e^{-m_{\rm eff} r}$ and applying the coupling condition $\alpha_t = \frac{3}{4}\alpha_s$ derived in Section 2, the relative magnitude of the residual potential compared to Newtonian gravity is bounded by $\alpha_s \mathcal{O}(v_{\rm typ}^4/c^4)$. This confirms the result of Section 2: even when the baryonic source is moving, the fifth force between baryons in 4D is suppressed, ensuring consistency with observations.

\subsection{Interaction Energy-Momentum Tensor on the 4D Brane}

We evaluate the interaction energy-momentum tensor on the 4D brane using the kinematic fields. Substituting the spatial fields into the scalar invariants yields $A_\mu A^\mu = g_v^2 (K^2 - V_i V_i)$ and $f_{\mu\nu}f^{\mu\nu} - f^2 \simeq \frac{2}{3} g_t^2 K^2 - 2g_t^2 V_i V_i$. The interaction Lagrangian evaluates to:
\begin{equation}
    \mathcal{L}_{\chi, {\rm int}} \simeq -\frac{1}{2}g_\chi \chi^2 \left[ K^2 \left( g_s^2 - g_v^2 + \frac{2}{3}g_t^2 \right) + V_i V_i \left( g_v^2 - 2g_t^2 \right) \right].
\end{equation}
Under the coupling ratio $g_s^2 : g_v^2 : g_t^2 = 4 : 6 : 3$ derived in Section 2, the relations $g_s^2 - g_v^2 + \frac{2}{3}g_t^2 = 0$ and $g_v^2 - 2g_t^2 = 0$ hold. Both the static and kinematic terms vanish ($\mathcal{L}_{\chi, {\rm int}} = 0$), preserving the zero mass shift ($\Delta m^2(x) = 0$).
The energy-momentum tensor $T_{\mu\nu}$ is derived from the metric variation. Since $\mathcal{L}_{\chi, {\rm int}} = 0$, the tensor is determined by the derivative term, $T_{\mu\nu} = g_\chi \chi^2 [ A_\mu A_\nu - 2 f_{\mu\lambda} f_\nu^{\;\lambda} + 2 f f_{\mu\nu} ]$. Note that in the non-relativistic limit ($v \ll c$), the $\mathcal{O}(v^2)$ kinematic corrections to the energy density are suppressed and can be neglected. Substituting the field components yields the energy density $T_{00}$:
\begin{equation}
    T_{00} \simeq g_\chi \chi^2 \left[ (-g_v K)^2 - 2 \left( \frac{4}{9}g_t^2 K^2 \right) + 2 \left(-\frac{1}{3}g_t K\right)\left(\frac{2}{3}g_t K\right) \right] = g_\chi \chi^2 K^2 \left( g_v^2 - \frac{4}{3}g_t^2 \right).
\end{equation}
Applying the relation $g_v^2 - \frac{4}{3}g_t^2 = \frac{1}{2}g_s^2$ and substituting $K(\vec{r})$, this reproduces the static energy density $T_{00} = \left[\frac{1}{2}\frac{g_\chi g_s^2}{(4\pi G)^2}\chi^2\right]\Phi_b^2(\vec{r}) e^{-2m_{\rm eff} r}$. The momentum density $T_{0i}$ evaluates to:
\begin{equation}
    T_{0i} \simeq g_\chi \chi^2 \left[ -g_v^2 K V_i - 2 \left( -\frac{1}{3}g_t^2 K V_i \right) + 2 \left(-\frac{1}{3}g_t K\right)(-g_t V_i) \right] = -g_\chi \chi^2 (K V_i) \left( g_v^2 - \frac{4}{3}g_t^2 \right).
\end{equation}
Finally, the spatial stress tensor $T_{ij}$ evaluates to:
\begin{equation}
\begin{split}
    T_{ij} &\simeq g_\chi \chi^2 \left[ g_v^2 V_i V_j - 2 \left( g_t^2 V_i V_j + \frac{1}{9}g_t^2 K^2 \eta_{ij} \right) + 2 \left(-\frac{1}{3}g_t K\right)\left(-\frac{1}{3}g_t K \eta_{ij}\right) \right] \\
    &= g_\chi \chi^2 \left( g_v^2 - 2g_t^2 \right) V_i V_j.
\end{split}
\end{equation}
Applying the coupling condition $g_v^2 = 2g_t^2$, the stress tensor vanishes ($T_{ij} = 0$). This confirms that the localized interaction energy behaves as a pressureless fluid, generating no internal spatial stress even when the baryonic source is in motion.

\end{document}